\documentclass[11pt,a4paper,english,superscriptaddress,nofootinbib,showpacs]{revtex4}
\usepackage{color}
\usepackage{amssymb}
\usepackage{amsmath}
\usepackage{graphicx}
\usepackage[T1]{fontenc}
\usepackage{hyperref}
\usepackage[brazil]{babel}
\usepackage[latin1]{inputenc}
\newcommand{\ba}{\begin{eqnarray}}
\newcommand{\ea}{\end{eqnarray}}
\newcommand{\be}{\begin{equation}}
\newcommand{\ee}{\end{equation}}

\usepackage{amsfonts}

\begin{document}

\title{Five-Dimensional $f(R)$ Braneworld Models}

\author{J. M. Hoff da Silva}
\email{hoff@feg.unesp.br, hoff@ift.unesp.br}\affiliation{UNESP - Campus de Guaratinguet\'a - DFQ, Av. Dr.
Ariberto Pereira da Cunha, 333 CEP 12516-410, Guaratinguet\'a-SP,
Brazil.}

\author{M. Dias}
\email{mafd@cern.ch}\affiliation{Departamento de Ci\^encias Exatas e da Terra,Universidade Federal de S\~ao Paulo\\
Diadema-SP-Brazil.}


\begin{abstract}
After incorporating the $f(R)$ gravity into the
general braneworld sum rules scope, it is shown that some particular class of
warped five-dimensional nonlinear braneworld models, which may be
interesting for the hierarchy problem solution, still require a negative
tension brane. For other classes of warp factors (suitable and not
suitable for approaching the hierarchy problem) it is not necessary
any negative brane tension in the compactification scheme. In this vein, it is argued that
in the bulk $f(R)$ gravity context, some types of warp factors may be useful for approaching the hierarchy problem
and for evading the necessity of a negative brane tension in the compactification scheme.
\end{abstract}
\maketitle

\section{Introduction}

The idea that our Universe is a submanifold (the brane) embedded
in a higher-dimensional spacetime led to a huge growth in the number
of new braneworld models being proposed. Brane gravity theories in
eleven dimensions were first introduced by Horava and Witten
\cite{HW}. In the Randall and Sundrum (RS) \cite{Randall1,
Randall2} braneworld models, the spacetime has five dimensions ---
an effective Horava-Witten scenario without the 6D Calabi-Yau
space --- endowed with a warped structure. In the so-called RSI
model \cite{Randall1} there are two 3-branes with opposite
tensions placed at the end of a $S^{1}/Z_{2}$ orbifold. One of the
branes has, typically, Planckian energy scales and naturally, via
the warp factor, the effective energy scale on the another brane
is about TeV.

The RSI model line element is given by \be
ds^{2}_{RSI}=e^{-2\kappa r_{c} |r|}\eta_{\mu\nu}dx^{\mu}dx^{\nu}+
r_{c}^{2}dr^{2}\label{j1},\ee where $\eta_{\mu\nu}$ is the usual
Minkowiski metric and $r_{c}$ is the compactification radius. The
Planck brane is placed at $r=0$, while the visible brane is placed at
$r=\pi$. Therefore, it is easy to verify that any mass parameter
on the visible brane will be dressed by the $e^{-2\kappa
r_{c}\pi}$ factor. Consequently, it is not necessary a large
hierarchy among the parameters in order to obtain the TeV scale
(the scale of the dressed masses) from the fundamental Planck
scale \cite{Randall1}.

In general, braneworld models are inspired in string theory and it
is expected that a considered model makes contact with some string
theory limits. In this vein, it is important to study braneworld
models in, for instance, modified Einstein-Hilbert gravity in the
hope that such a bottom-up approach turns out to be useful from
the phenomenological point of view. Here we investigate how a
$f(R)$ gravity theory in the bulk may act in the braneworld
context (for recent reviews in $f(R)$ gravity see \cite{revfR}).
Higher derivative gravity in the braneworld framework was analyzed
in several contexts. For instance, in Ref. \cite{mb} it is shown how $f(R)$
theories may evade the fine-tuning problem present in the RSI
model while cosmological aspects of $f(R)$ braneworld gravity were
studied in Ref. \cite{am1} (see also \cite{am2,Pe}).

In this work we are concerned with the following issue:  the RSI model has one negative brane tension in the
compactification scheme. It is indeed necessary in the model. However, a negative brane tension is not a gravitationally stable object by itself. Let us discuss this issue a little further. If it is allowed to the branes
to vary their position (which is not the case in the RSI model), then it is possible to
show that the resulting RSI-like scenario is free from gravitational instability, since the radion is massless
and may be made heavy via the mechanism presented in Ref. \cite{GW}. By the same token, if the
distance between the branes may vary, it is possible to show that the effective gravitational constant
on the brane do not depends on the brane tension signal. However, if the distance between the branes is fixed a priori (i. e. the radion field is frozen), as in the RSI case, then the effective Newtonian constant on the brane depends linearly on the brane tension,
leading to an ill defined scenario for negative tension branes. Necessary conditions for circumventing the
necessity of negative brane tensions were found in the context of
3-branes embedded into a five-dimensional bulk respecting
Brans-Dicke gravity \cite{enois}. Roughly speaking, the presence
of terms proportional to the scalar field, $\phi$, (and their derivatives)
relaxes the constraints imposed by the consistency conditions,
allowing a braneworld setup without the necessity of a
negative brane tension in the compactification scheme. In this
vein, one may hope that, at least in some regime, the $f(R)$ bulk
gravity acts in the same way. In fact, keeping in mind the
conformal equivalence between $f(R)$ and scalar-tensorial theories
\cite{revfR}, the use of $f(R)$ in order to avoid a negative brane
tension is expected to work. Note, however, that this equivalence requires a nonzero potential for the scalar field and in Ref. \cite{enois} $V(\phi)=0$. Therefore, this work concerns essentially to a new case. Here, we show that for some classes of warp factors the previous considerations are correct, i. e.,
$f(R)$ theories indeed help to evade the necessity of a
negative brane tension even without to varying the distance between the branes. However, for another class of warp factors (among them the usually used for the hierarchy problem solution)
the scenario including $f(R)$ theory in the bulk still needs a
negative brane tension.

This paper is structured as follows: In the next Section we start from a short review on the consistency conditions
obtained for the usual General Relativity case, making explicit the
necessity of a negative brane tension in the RSI framework. Then we move forward and generalize the consistency conditions to the bulk $f(R)$ gravity case. In Section III we conclude with a final discussion.

\section{The $f(R)$ braneworld models}

Here we shall first reobtain, for completeness, the braneworld sum rules
in General Relativity. Gibbons, Kallosh and Linde \cite{Gibbons}
prompted to define a set of simple rules aimed at checking the
consistency of five-dimensional models case by case. After that, a
work of Leblond, Myers and Winters \cite{Leblond} extended these
{\it sum rules} for higher dimensions. We shall reproduce their
arguments bellow.

We start analyzing a $D$-dimensional bulk spacetime endowed with a non-factorable geometry, whose metric is given by
\[ds^2=G_{AB}dx^Adx^B=W^2(r)g_{\mu\nu}dx^\mu dx^\nu+g_{ab}(r) dr^a dr^b,\]
where $W^2(r)$ is the warp factor, $x^A$ denotes the coordinates
of the full $D$-dimensional spacetime, $x^\mu$ stands for the $(p +
1)$ non-compact coordinates of the spacetime and $r^a$ labels the
$(D-p-1)$ directions in the internal compact space. Note that this
type of metric encodes the possibility of existing q-branes $(q >
p)$, so that the $(q - p)$ extra dimensions are compactified on
the brane and constitute part of the internal space. As an
example, if $D=5, p=3$ and $W(r)=e^{-2kr_{c}|r|}$ one arrives at the RS
model.

The $D$-dimensional Einstein equations defined in the bulk are
\begin{equation}
\label{aldo1}R_{AB}=8\pi G_D\left(T_{AB}-\frac{1}{D-2}G_{AB}T^M_M\right),
\end{equation}
where  $G_D$ is the $D$-dimensional gravitational constant. We decompose the traced version of Ricci tensor components
\begin{equation}
\label{aldo2}R^\mu_\mu=\frac{8\pi G_D}{D-2}\left[(D-p-3)T^\mu_\mu-(p+1)T^a_a\right],
\end{equation}
where $T^\mu_\mu\equiv W^{-2}g^{\mu\nu}T_{\mu\nu}$ and
\begin{equation}
\label{aldo3} R^a_a=\frac{8\pi G_D}{D-2}\left[(p-1)T^a_a-(D-p-1)T^\mu_\mu\right],
\end{equation}
both written in terms of the bulk stress-energy tensor.

Furthermore, the D-dimensional spacetime Ricci tensor can be related with the brane Ricci tensor and with the warp factor $W$ by the equations
\begin{equation}\label{1}
R_{\mu\nu} = \bar{R}_{\mu\nu} -
\,\frac{g_{\mu\nu}}{(p+1)W^{p-1}}\nabla^2W^{p+1}\end{equation} and
\begin{equation}\label{2}R_{ab} = \tilde{R}_{ab} - \,\frac{p+1}{W} \nabla_a\nabla_b W,\end{equation}
where $R_{ab}$ , $\nabla_a$ and $\nabla^2$ are respectively the
Ricci tensor, the covariant derivative and the Laplacian operator
constructed by the internal space metric $g_{ab}$ . $\bar{R}_{\mu\nu}$
is the Ricci tensor derived from $g_{\mu\nu}$ . The three
curvature scalars are denoted by $R = G^{AB}R_{AB},
\bar{R}=g^{\mu\nu} \bar{R}_{\mu\nu}$ and $\tilde{R}=g^{ab}\tilde{R}_{ab}$.
Thus, the traces of Eqs. (\ref{1}) and (\ref{2}) yield to the
following equations
\begin{equation}
\label{3}\frac{1}{p+1}\left(W^{-2}\bar{R}-R^\mu_\mu\right)=pW^{-2}\nabla W\cdot \nabla W+W^{-1}\nabla^2W
\end{equation}
and
\begin{equation}
\label{4}\frac{1}{p+1}\left(\tilde{R}-R^a_a\right)=W^{-1}\nabla^2W,
\end{equation}
where $R^\mu_\mu\equiv W^{-2} g^{\mu\nu}R_{\mu\nu}$ and
$R^a_a\equiv g^{ab}R_{ab}$ (related by $R=R^\mu_\mu+R^a_a$).
Supposing that $\alpha$ is an arbitrary constant, it is not
difficult to see that
\begin{equation}
\label{5}\nabla\cdot(W^\alpha\nabla W)=W^{\alpha+1}(\alpha W^{-2}\nabla W\cdot\nabla W+W^{-1}\nabla^2W).
\end{equation}
Combining Eqs. (\ref{3}), (\ref{4}) and (\ref{5}) will allow us to write
\begin{equation}
\label{6}\nabla\cdot(W^\alpha\nabla
W)=\frac{W^{\alpha+1}}{p(p+1)}\left[\alpha\left(W^{-2}\bar{R}-R^\mu_\mu\right)+(p-\alpha)\left(\tilde{R}-R^a_a\right)\right].
\end{equation}

As an {\it ansatz} for the stress-energy tensor we can write
\begin{equation}
\label{7}T_{AB}=-\Lambda G_{AB}-\sum_iT^{(i)}_qP\left[G_{AB}\right]_q^{(i)}\Delta ^{(D-q-1)}(r-r_i)+\tau_{AB},
\end{equation}
where $\Lambda$ is the bulk cosmological constant, $T_q^{(i)}$ is
the $\textnormal{i}^{\textnormal{th}}$ q-brane tension, $\Delta
^{(D-q-1)}(r-r_i)$ is the covariant combination of delta functions
which localizes the brane, $P\left[G_{AB}\right]_q^{(i)}$ is the
pull-back of the spacetime metric to the worldvolume  of the
$q$-brane (or the induced metric on the brane) and any other
matter contribution  is represented by $\tau_{AB}$. Following this
{\it ansatz} one obtains
\begin{equation}
\label{8} T^\mu_\mu = -(p + 1)\Lambda + \tau_\mu^\mu - \sum_i T_q^i \Delta^{(D-q-1)} (r - r_i )(p + 1)
\end{equation}
and
\begin{equation}
\label{9}T_a^a = -(D-p-1)\Lambda + \tau_a^a - \sum_i T_q^i \Delta^{(D-q-1)} (r - r_i )(q - p).
\end{equation}

Inserting Eqs. (\ref{aldo1}) and (\ref{aldo2}) in Eq. (\ref{6}), with
$T^\mu_\mu$ and $T^a_a$ as defined in Eqs. (\ref{8}) and (\ref{9})
respectively, it is straightforward to demonstrate --- following
the previous {\it ansatz} (\ref{8},\ref{9}) and performing an
integration in the compact internal space as well --- that
\begin{eqnarray}\nonumber
&&\label{10}\oint W^{\alpha+1}\left\{\alpha \bar{R}W^{-2}+(p-\alpha)\tilde{R}-\left[\frac{p+1}{D-2}[(p-2\alpha)(D-p-1)+2\alpha]+\frac{(D-p-1)p(2\alpha-p+1)}{D-2}\right]\Lambda\right.\nonumber\\
&&\left.-8\pi G_D\left[\sum_i\left[\frac{p+1}{D-2}((p-2\alpha)(D-p-1)+2\alpha)+(q-p)\frac{p(2\alpha-p+1)}{D-2}\right]T_q^{(i)}\Delta^{(D-q-1)(r-r_i)}\right.\right.\nonumber\\
&&\left.\left.-\frac{(p-2\alpha)(D-p-1)+2\alpha}{D-2}\tau^\mu_\mu-\frac{p(2\alpha-p+1)}{D-2}\tau^a_a\right]\right\}=0.
\end{eqnarray}

We shall refer to Eq. (\ref{10}) as the one parameter family of
consistency conditions for brane world scenarios. If one consider,
as an example, the case $D=5, p=3$, with $\alpha=-1$, setting
$\tau^a_a=\tau^\mu_\mu=0$ the equation above reduces simply to
 \begin{equation}
-\bar{R}\oint W^{-2}=32\pi G_5\sum_iT_3^{(i)}\label{PULTIMA},
\end{equation}
since $\tilde{R}=0$ for a single internal direction.

In trying to reproduce our Universe, it is acceptable to set
$\bar{R}=0$ and Eq. (\ref{PULTIMA}) reduces to
\begin{equation}
32\pi G_5\sum_iT_3^{(i)}=0\label{ULTIMA},
\end{equation} making necessary the presence of a negative brane
tension in the Randall-Sundrum setup.

After this short review, let us start the analysis of the $f(R)$ case itself by considering the nonlinear (in the Ricci scalar) bulk action in $D$ dimensions given by

\be S_{bulk}=-\frac{1}{8\pi G_{D}}\int d^{D}x
\sqrt{-G}[f(R)+\mathcal{L}_{m}],\label{j2}\ee where $\mathcal{L}_{m}$ stands for the matter (and brane) lagrangian. The variation of the above action with respect to the bulk metric gives the follow field equation

\be F(R)R_{AB}-\frac{1}{2}G_{AB}f(R)+G_{AB}\Box F(R)-\nabla_{A}\nabla_{B}F(R)=8\pi G_{D}T_{AB},\label{j3}\ee which may be recast in the more familiar form \be
R_{AB}-\frac{1}{2}RG_{AB}=\frac{1}{F(R)}\Bigg[8 \pi
G_{D}T_{AB}-\Bigg(\frac{1}{2}RF(R)-\frac{1}{2}f(R)+\Box
F(R)\Bigg)G_{AB}+\nabla_{A}\nabla_{B}F(R)\Bigg],\label{j4}\ee
being $F(R)=df(R)/dR$. In order to implement the bulk
$f(R)$ gravity in the consistency conditions program we call attention to the fact that Eq. (\ref{6}) is
obtained without any reference to the field equation in question,
being instead purely geometric. Therefore, it is a suitable
starting point to our generalization. From Eq. (\ref{j4}) it is
easy to see that the Ricci scalar is given by \be
R=\frac{2}{(2-D)F(R)}\Bigg[8\pi
G_{D}T-\frac{D}{2}RF(R)+\frac{D}{2}f(R)+(1-D)\Box
F(R)\Bigg].\label{j5}\ee Note that, obviously, it is possible to
isolate the Ricci scalar obtaining $R=[8\pi
G_{D}T+(D/2)f(R)+(1-D)\Box F(R)]/F(R)$. However, we shall develop
our presentation with Eq. (\ref{j5}) for the sake of clarity, since in this way the usual
General Relativity limit ($f(R)=R$ and $F(R)=1$) is easily
obtained. From Eq. (\ref{j5}) the contact with General Relativity
is, indeed, more explicit.

Substituting Eq. (\ref{j5}) in (\ref{j4}) we have\be
R_{AB}=\frac{1}{F(R)}\Bigg[8\pi
G_{D}\Bigg(T_{AB}-\frac{G_{AB}}{D-2}T\Bigg)+\nabla_{A}\nabla_{B}F(R)+\frac{1}{D-2}G_{AB}[RF(R)-f(R)+\Box
F(R)] \Bigg],\label{j6}\ee whose limit to General Relativity gives
only the first term, as expected. The partial traces of the above
equation, following the same standard previous notation,
are given by \ba
R^{\mu}_{\mu}&=&\left.\frac{1}{(D-2)F(R)}\Bigg[8\pi
G_{D}[(D-p-3)T^{\mu}_{\mu}-(p+1)T^{m}_{m}]+(D+p-1)W^{-2}\nabla^{\mu}\nabla_{\mu}F(R)
\right.\nonumber\\&+&\left.(p+1)[RF(R)-f(R)+\nabla^{m}\nabla_{m}F(R)]
\Bigg]\right.\label{j7} \ea and

\ba R^{m}_{m}&=&\left.\frac{1}{(D-2)F(R)}\Bigg[8\pi
G_{D}[(p-1)T^{m}_{m}-(D-p-1)T^{\mu}_{\mu}]+
(2D-p-3)\nabla^{m}\nabla_{m}F(R)\right.\nonumber\\&+&\left.(D-p-1)[RF(R)-f(R)+W^{-2}\nabla^{\mu}\nabla_{\mu}F(R)]\Bigg].\right.\label{j8}\ea

Now it is possible to adequate Eq. (\ref{6}) to the $f(R)$ bulk
gravity case. Hence with Eqs. (\ref{j7}) and (\ref{j8}) it reads
\ba \nabla\cdot(W^{\alpha}\nabla
W)&=&\left.\frac{W^{\alpha+1}}{p(p+1)(D-2)F(R)}\Bigg[(D-2)F(R)[\alpha
\bar{R}W^{-2}+(p-\alpha)\tilde{R}]+8\pi
G_{D}T^{\mu}_{\mu}[(p-\alpha)\right.\nonumber\\&\times&\left.(D-p-1)-\alpha(D-p-3)]+8\pi
G_{D}T^{m}_{m}[\alpha(p+1)-(p-\alpha)(p-1)]\right.\nonumber\\&-&\left.
[RF(R)-f(R)][\alpha(p+1)+(D-p-1)(p-\alpha)]-W^{-2}\nabla_{\mu}\nabla^{\mu}F(R)[(D-p-1)
\right.\nonumber\\&+&\left.\alpha(D+p-1)]-\nabla^{m}\nabla_{m}F(R)[\alpha(p+1)+(p-\alpha)(2D-p-3)]
\Bigg].\label{j9}\right. \ea In an internal compact space the integral of the equation above vanishes, since $\oint \nabla \cdot(W^{\alpha}\nabla W)=0$. In order to complete the analysis we shall use the comprehensive stress-tensor ansatz defined by Eq. (\ref{7}) with the partial traces given by Eqs. (\ref{8}) and (\ref{9}). Substituting Eqs. (\ref{8}) and (\ref{9}) into Eq. (\ref{j9}) we arrive, after some algebra, at the generalization of Eq. (\ref{10}) for the $f(R)$ bulk gravity case:
\ba && \left. \oint \frac{W^{\alpha+1}}{F(R)}\Bigg\{\Big[(D-2)F(R)[\alpha \bar{R}W^{-2}+(p-\alpha)\tilde{R}] \Big]- [RF(R)-f(R)+2\Lambda][(D-p-1)(p-\alpha)\right.\nonumber\\&+&\left.\alpha(p+1)]-8\pi G_{D}\Big\{(p+1)[(D-p-1)(p-2\alpha)+2\alpha]+p(1+2\alpha-p)(q-p)\Big\}\sum_{i}T_{q}^{i}\Delta^{(D-q-1)}(r-r_{i})
\right.\nonumber\\&+&\left.8\pi G_{D}[(D-p-1)(p-2\alpha)+2\alpha]\tau^{\mu}_{\mu}+8\pi G_{D}p(1+2\alpha-p)\tau^{m}_{m}-W^{-2}\nabla_{\mu}\nabla^{\mu}F(R)\Big[(D-1)(\alpha+1)\right.\nonumber\\&+&
\left.p(\alpha-1)\Big]-\nabla_{m}\nabla^{m}F(R)\Big[\alpha(p+1)+(p-\alpha)(2D-p-3)\Big]\Bigg\}=0.\right.\label{j10}\ea

In order to make our point clearer, we shall particularize the analysis to the five-dimensional bulk case. Therefore, we take $D=5$, $p=q=3$, and for simplicity we set $\tau^{\mu}_{\mu}=\tau^{m}_{m}=0$, disregarding any extra bulk matter contribution. Note that for this case $\tilde{R}=0$, since there is only one extra (transverse) dimension. With such particularizations Eq. (\ref{j10}) reads
\ba && \left. \oint \frac{W^{\alpha+1}}{F(R)}\Bigg\{\alpha F(R)\bar{R}W^{-2}-(\alpha+1)[RF(R)-f(R)+2\Lambda]-32\pi G_{5}\sum_{i}T_{3}^{i}\delta(r-r_{i})\right.\nonumber\\&-&\left.\frac{W^{-2}}{3}(1+7\alpha)\nabla_{\mu}\nabla^{\mu}F(R)-4\nabla^{m}\nabla_{m}F(R) \Bigg\}=0.\right.\label{j11} \ea In trying to describe our Universe, it is conceivable to set $\bar{R}$ which vanishes with an accuracy of $10^{-120}M_{Planck}^{2}$. Besides, from Eqs. (\ref{aldo2}) and (\ref{aldo3}) it is easy to see that the bulk scalar of curvature for the case in question is given only in terms of the warp factor by
\be R=R^{\mu}_{\mu}+R^{m}_{m}=-\frac{4}{W}\nabla^{2}W-\frac{1}{W^{4}}\nabla^{2}W^{4},\label{j12}\ee where $\nabla^{2}=\frac{1}{2g_{rr}}(\partial_{r}g_{rr})g^{rr}\partial_{r}+(\partial_{r}g^{rr})\partial_{r}+g^{rr}
\partial^{2}_{r}$. For the simplest case of stabilized distance between the branes ($g_{rr}$ constant) --- as in the RSI model --- we have simply $\nabla^{2}=\partial_{r}^{2}$, where $g_{rr}=1$ for simplicity. In view of Eq. (\ref{j12}), and taking into account that the warp factor depends only on the extra dimension, we have
\ba && \left. \oint \frac{W^{\alpha+1}}{F(W)}\Bigg\{\frac{(\alpha+1)}{4}\Bigg[2\Lambda-f(W)-F(W)\Bigg(\frac{4}{W}\nabla^{2}W+
\frac{1}{W^{4}}\nabla^{2}W^{4}\Bigg)\Bigg]\right.\nonumber\\&+&\left.8\pi G_{5}\sum_{i}T_{3}^{i}\delta(r-r_{i})+\nabla^{2}F(W)\Bigg\}=0.\right.\label{j13}\ea It is important to stress that for $f(R)=R=-\frac{4}{W}\nabla^{2}W-\frac{1}{W^{4}}\nabla^{2}W^{4}$ the usual sum rules for General Relativity are recovered.

The most direct way to obtain a necessary condition concerning the brane tension sign is choosing $\alpha=-1$ \cite{Gibbons,Leblond}, since it eliminates the overall warp factor and the first term of Eq. (\ref{j13}). Hence, in this particulary interesting case we have \be \oint \frac{1}{F(W)}\Big\{8\pi G_{5}\sum_{i}T_{3}^{i}\delta(r-r_{i})+\nabla^{2}F(W)\Big\}=0,\label{j14} \ee which leads to the following necessary condition \be 8\pi G_{5}\sum_{i}\frac{T_{3}^{i}}{F_{i}(W)}+\oint\frac{\nabla^{2}F(W)}{F(W)}=0,\label{j15}\ee where $F_{i}(W)=F(W(r=r_{i}))$. Note that, according to Eq. (\ref{j15}), a negative brane tension in the compactification scheme is not necessary, provided that\footnote{Note that the warp factor must be of class $C^{4}$ at least.} $\oint\frac{\nabla^{2}F(W)}{F(W)}<0$. We remark by passage that the claim above is not applicable in the pathological case given by
\be \nabla^{2}F(W)+\frac{(\alpha+1)}{4}\Bigg[2\Lambda -f(W)-F(W)\Bigg(\frac{4}{W}\nabla^{2}W+
\frac{1}{W^{4}}\nabla^{2}W^{4}\Bigg)\Bigg]=0,\label{j16}\ee which, certainly, is not the most general case.

As it is possible to evade the necessity of a negative brane tension by means of a condition which depends on the warp factor, it would be interesting to explore a little further the analysis studying the possibility of warp factors being both suitable for the hierarchy problem solution and satisfying $\oint\frac{\nabla^{2}F(W)}{F(W)}<0$ at the same time.

In five dimensions it is easy to see that the bulk scalar of curvature is given by \be R=-\frac{12(W^{'}(r))^{2}+8W(r)W^{''}(r)}{(W(r))^{2}},\label{s1}\ee where a comma means derivative with respect to $r$. Obviously, any warp factor leading to a constant $R$ --- even solving the hierarchy problem --- does not satisfy the constraint given by Eq. (\ref{j15}), and shall not be used to set a compactification scheme without a negative brane tension.

There are several warp factors profiles which may be used in order to approach the hierarchy problem. To fix ideas, let us work with the simplest choice $W(r)=e^{-ar^{n}}$, being $a(>0)$ and $n$ constants\footnote{We shall not be concerned with the absolute value of $r$ in this simple example.}. From Eq. (\ref{s1}) we have \be R=-4nar^{n-2}\Big[5nar^{n}-2(n-1)\Big],\label{s2}\ee hence models with $n=1$ should be excluded. Following this reasoning, it is not difficult to test a bulk $f(R)$ gravity braneworld model as a candidate to approach the hierarchy problem without a negative brane tension. In our simple example outlined above, if $n\neq 1$ then in principle it would be possible to use such a warp factor to attack both problems. Of course, to complete the analysis one should to fix the specific model by setting the functional form of $f(R)$.

\section{Concluding Remarks}

The generalization of the braneworld consistency conditions to the $f(R)$ bulk gravity was considered in order to derive criteria on the possible warp factors which could solve the hierarchy problem leading, at the same time, to a consistent compactification framework. By applying this program it is possible to verify, among several brane world scenarios, what type of models may be phenomenologically useful.

By inspecting Eq. (\ref{j15}) it is simple to see that any braneworld model whose warp factor leads to a constant bulk scalar of curvature necessarily requires at least one negative brane tension. For some typical warp factors (suitable for the hierarchy problem solution) the constraint (\ref{j15}) can not be satisfied.

On the other hand, if the model does not concern with the hierarchy problem --- leaving it for supersymmetric extensions of the standard model, for instance --- then Eq. (\ref{j15}) is the relevant constraint. A non-constant $R$ (in respect to the extra transverse dimension) may lead to a scenario without a negative brane tension. The presence of $F(R)$ terms in the consistency conditions acts in order to relax the conditions to be satisfied, opening the possibility for a well defined model. It is certainly a gain when compared with usual General Relativity based braneworld models.

It is important to remark that in the bulk $f(R)$ gravity context, other type of warp factors may be useful for approach the hierarchy problem and to evade the necessity of a negative brane tension in the compactification scheme. In this approach the $F(R)$ (and accordingly the $F(W)$) term just depends on the extra transverse dimension, in such a way that it is not expected any experimental contradiction with local tests of gravity on the brane. However, in order to give a cogent argument one should to project the $f(R)$ bulk geometrical quantities on the brane, via e. g. the Gauss-Codazzi formalism with appropriated junction conditions and calculate the relevant post Newtonian parameters. On the other hand, keeping in mind the conformal equivalence between $f(R)$ and scalar tensor theories, one may speculate that, indeed, there is no violation of any experimental bound, i. e., Eq. (\ref{j15}) is in fact well defined from the phenomenological point of view.

\end{document}